\documentclass[twocolumn,prl,aps,showpacs]{revtex4}

\usepackage{graphicx}
\usepackage{dcolumn}
\usepackage{bm}

\begin{document}

\preprint{APS/123-QED}

\title{Quantitative and Conceptual Considerations for Extracting the Knudsen Number in Heavy Ion Collisions}

\author{J.L. Nagle}
 \affiliation{University of Colorado, Boulder, CO}
 \email{Jamie.Nagle@Colorado.Edu}
\author{P. Steinberg}
\affiliation{Brookhaven National Laboratory, Upton, NY}
\author{W.A. Zajc}
 \affiliation{Columbia University, New York, NY}

\date{\today}
\begin{abstract}
%
In this paper we examine the methodology for extracting the Knudsen number ($K$) and the ratio of
shear viscosity to entropy density ($\eta/s$) developed by Drescher {\it et al}~\cite{Drescher:2007cd}.
The final result for $\eta/s$ turns out to be quite sensitive to Glauber parameters, and particularly the
parameter $x$ which controls the balance between $N_{part}$ and $N_{coll}$.  
We also explore how alternative formulations of the functional relation between the elliptic flow and Knudsen
number ($K = \lambda / \overline{R}$) impacts the physics conclusions, based on Pad\'e
approximants.
Finally, we extend the
calculation to include a limiting minimum value on the mean free path proportional to the DeBroglie wavelength.
These results emphasize the importance of clarifying the initial state used in different calculations, as well as the ambiguities inherent in using a transport approach in a strongly-coupled regime.

\end{abstract}

\maketitle

\section{Introduction}

The medium created in heavy ion collisions at the Relativistic Heavy
Ion Collider (RHIC) defied expectations by showing a strong
collective flow, characteristic of a perfect fluid~\cite{Arsene:2004fa,Adcox:2004mh,Back:2004je,Adams:2005dq}. This
behavior is particularly striking since even a modest amount of
viscous damping, parametrized as the ratio of shear viscosity
$\eta$ to the entropy density $s$,  is predicted to result in large
deviations from ideal hydrodynamics~\cite{Teaney:2003kp}. Several methods, each subject
to as-yet uncontrolled systematic uncertainties, have been developed
to estimate the value of ($\eta/s$) from the experimental data on
collective flow~\cite{Lacey:2006bc,Drescher:2007cd},
fluctuations~\cite{Gavin:2006xd}, and heavy quark
transport~\cite{Adare:2006nq}.
%
%
A parallel effort in this direction has been the development of hydrodynamic codes
in two and three dimensions
with a properly causal and stable relativistic treatment of viscous effects to second order
in the velocity gradients~\cite{Song:2007ux,Romatschke:2007mq,Luzum:2008cw,Molnar:2008xj,Chaudhuri:2008sj}.
While recent work has resulted in a greatly improved understanding
of the various formalisms used by these authors~\cite{TECHQM},
and direct comparison to experimental data has been made for a range
of values in $\eta/s$, there remains considerable
debate regarding the sensitivity to initial
conditions, equations of state, and precise freeze-out conditions.

These different approaches are in large part motivated by the
conjecture~\cite{Kovtun:2004de} that there is a fundamental bound on the viscosity to
entropy density ratio: $\eta/s \geq 1/4\pi$. Intriguingly, all methods produce
values for $\eta/s$ from the RHIC fluid that are within factors of 1-4 of the bound.
This brings into sharp focus the need to understand in detail the utility and the
limitations of each method. In this paper we concentrate on the approach of
Drescher {\em et al.}~\cite{Drescher:2007cd}, which provides a useful framework for parameterizing
departures from ideal hydrodynamic behavior in terms of the Knudsen number
$K \equiv \lambda/\ \bar R$, where $\lambda$ is the mean free path of particles or quasi-particles in the system
and $\bar R$ is a characteristic
measure of the system size.

It has been realized since the earliest
applications of hydrodynamics to nuclear collisions (see the first two footnotes of Ref.~\cite{Belenkij:1956cd})
that the determination of $K$ is directly related to the viscosity of the fluid.
For sufficiently small values of $K$ the hydrodynamic limit is reached,
but larger values imply the presence of a non-trivial mean free path,
leading to the damping of momentum anisotropies, and thus deviations from ideal fluid behavior.
In Ref.~\cite{Drescher:2007cd} the Knudsen number is directly related to the
elliptic flow scaled by the initial state eccentricity $v_2/\epsilon$, thereby
providing a convenient link between the experimental data and the transport properties of the medium.

\section{Methodology}

In this section we review the previous extraction of the Knudsen number and the
the ratio $\eta/s$ as formulated in Ref.~\cite{Drescher:2007cd}.
In doing so, we demonstrate the sensitivity to various assumptions at different steps in the calculation.
Following Bhalerao {\it et al.}~\cite{Bhalerao:2005mm}, Ref.~\cite{Drescher:2007cd}
takes the characteristic size  of the system $\overline{R}$
as the scale of the strongest gradient in the initial matter
configuration of the system, estimated by
\begin{equation}
{\overline{R}}= \frac{1}{\sqrt{1/\langle x^2 \rangle + 1/\langle y^2 \rangle }}
\end{equation}
\noindent
This expression for $\overline{R}$ is used to determine the Knudsen number,
which in turn is assumed to describe deviations of the
the elliptic flow parameter ($v_2$) scaled by the
initial eccentricity $\epsilon$ from the corresponding ratio for
ideal hydrodynamics $(v_2/\epsilon)_{ih}$ via
\begin{equation}
\frac{v_{2}}{\epsilon} = \left( \frac{v_{2}}{\epsilon} \right) _{ih} {{1} \over {1 + K/K_{0}}}
\label{eqn_drescher}
\end{equation}
where $K_{0}$ is a parameter estimated to be approximately 0.7 from varying the particle cross-section
in a 3D transport code~\cite{Chen:2005mr}, or by comparison to a Monte Carlo solution of a two-dimensional Boltzmann
equation~\cite{Gombeaud:2007ub}.  The parametric dependence on $K$ in Eqn.~\ref{eqn_drescher}
is not directly motivated  by an underlying microscopic theory,
but rather was proposed in Ref.~\cite{Bhalerao:2005mm} to have 
the correct limits in the two extremes of large and small $K$.
In the limit of large mean-free path  (large $K$)
${v_{2}}/{\epsilon} \propto 1/K$,
while in the small mean-free path (small $K$) limit ${v_{2}}/{\epsilon}$ approaches the
limiting ideal fluid or hydrodynamic value $(v_{2}/\epsilon)_{ih}$ with corrections linear in $K$.
If the functional form of Eqn.~\ref{eqn_drescher} was unique (to be discussed below),
then it is clear that by combining the value of  $(\frac{v_{2}}{\epsilon})_{ih}$
calculated using ideal hydrodynamics with an experimental measure
of $v_2/\epsilon$ would permit a direct determination of $K$
(albeit with the embedded uncertainties of hydrodynamic calculations mentioned in the introduction).
Even more appealing is the possibility developed in Ref.~\cite{Drescher:2007cd}
of directly determining {\em both} $(\frac{v_{2}}{\epsilon})_{ih}$ and $K$ by
fitting the experimental data on $v_2/\epsilon$ as a function of centrality.

The Knudsen number varies with centrality both directly via associated changes
in $\overline{R}$ and indirectly through changes in the particle number density $n$:
\begin{equation}
\label{eqn_Kn}
\frac{1}{K} = \frac{\overline{R}}{\lambda} = \overline{R} n \sigma
\end{equation}
where $\sigma$ is the effective inter-particle cross-section.
Please note that having well-defined values for 
$n$, $\sigma$ and $\lambda$ implicitly assumes classical ballistic transport.
The prescription for estimating $n$ from the experimental data comes
from Ref.~\cite{Bhalerao:2005mm}
\begin{equation}
n =  \frac{1}{ S_{T} \tau}\  \frac{dN}{dy}
\end{equation}
which relies on calculations (for example from Monte Carlo Glauber~\cite{Alver:2008aq})
to determine the transverse area $S_T$ of the system
and the assumption of Bjorken expansion\cite{Bjorken:1982qr}
with $\tau$ as an appropriate proper time in a rapidity slice.
The characteristic time over which the gradients develop
was assumed in Ref.~~\cite{Bhalerao:2005mm} to be
$\tau = \overline{R}/c_s$ where $c_s$ is the speed of sound from the equation of state,
giving:
\begin{equation}
n(\tau = \overline{R}/c_s) =  \frac{1}{S_T}\  \frac{c_s}{\overline{R}}\ \frac{dN}{dy}
\label{eqndensity}
\end{equation}
and thus (by Eqn.~\ref{eqn_Kn}):
\begin{equation}
\frac{1}{K} =  \frac{\sigma c_{s}}{S_{T}} \frac{dN}{dy}
\end{equation}
Note that with these assumptions the Knudsen number
no longer requires knowledge of $\overline{R}$.
One can then eliminate the Knudsen Number entirely and write Eqn.~\ref{eqn_drescher} in terms of these
new quantities.
\begin{equation}
\frac{v_{2}}{\epsilon} = \left[ \frac{v_{2}}{\epsilon} \right]_{ih} {{1} \over {1 + {{1} \over {K_{0}(\sigma c_{s})(\frac{1}{S_{T}} \frac{dN}{dy})}}}}
\label{eqn_drescher2}
\end{equation}
which indicates that fitting plots of the experimental values of $v_2/\epsilon$ as a function 
of another `experimental quantity $\frac{1}{S_T} \frac{dN}{dy}$ can be used to determine
$(\sigma c_s$) and $(\frac{v_{2}}{\epsilon})_{ih}$.
This approach is then applied to PHOBOS measurements of  $v_2$ in Cu+Cu and Au+Au collisions at $\sqrt{s_{NN}} = 200$ GeV for
unidentified hadrons at pseudo-rapidity $\eta=0$
as a function of the number of participating nucleons ($N_{part}$)~\cite{Back:2004mh,Alver:2006wh,Alver:2008zza}.
We note here and below that the $dN/dy$ that appears in these equations is that for
total particle (presumably parton) rapidity density, so that further (plausible) assumptions are required
to express this in terms of the experimentally measured charged-particle
pseudo-rapidity density $dN_{ch}/d\eta$.

\section{Details of the Input Parameters}

In this section we describe the uncertainties
associated with the determination of the eccentricity-scaled elliptic flow
$v_2/\epsilon$ and the parton transverse areal density $\frac{1}{S_T} \frac{dN}{dy}$.
Close examination finds that the PHOBOS data presented in
Fig.~1 of Ref.~\cite{Drescher:2007cd} appears quite different from what is
nominally the {\it same} data presented in Fig.~6 of Ref.~\cite{Alver:2006wh}.
This is explained primarily by the fact that while the extraction of the Knudsen number
is based on experimental data, the approach relies on
various geometric quantities (e.g. the area $S_T$, the eccentricity $\epsilon$)
which are not measured directly, but which are only estimated for each centrality bin.
Detailing the differences is instructive and points out systematic uncertainties that
must be addressed before a fully quantitative estimate of $K$ can be made.

One significant source of uncertainty in all such calculations is the lack of knowledge
about the  initial distribution of energy and matter that is relevant for
the calculation of the initial eccentricity $\epsilon$ and the overlap area $S_{T}$.
In Ref.~\cite{Drescher:2007cd}, two initial conditions are presented representing
different assumptions about the initial state,
one from a Monte Carlo Glauber (MCG) calculation and the other from a particular Color Glass Condensate (CGC)
calculation~\cite{Drescher:2006ca}.
For the MCG case the initial distribution of energy is defined by a combination of 
spatial coordinates of participating nucleons and binary collisions.  The relative weights of 
the contributions follow the ``two-component'' model of Kharzeev and Nardi\cite{Kharzeev:2000ph}:
\begin{equation}
\frac{dN_{ch}}{d\eta} = n_{pp}\left[ (1 - x) \frac{N_{part}}{2} + x N_{coll} \right]
\label{eqn_twocomponent}
\end{equation}
with $x=0.20$, the so-called ``80:20'' mixture.
This nomenclature is potentially misleading: In central Au+Au collisions
$N_{coll}\sim 6N_{part}$, which implies that the $N_{part}$ term contributes only 40\% of the matter while
$N_{coll}$ contributes 60\%.  It is also notable that the chosen value of $x = 0.20$
is not commonly used in the literature (e.g. Ref.~\cite{Kharzeev:2000ph}) and the experimental
data at mid-rapidity at all beam energies is best described by 
$x = 0.13 \pm 0.01 (stat) \pm 0.05 (sys)$~\cite{Back:2004dy}.

\begin{figure}
\includegraphics[width=\columnwidth]{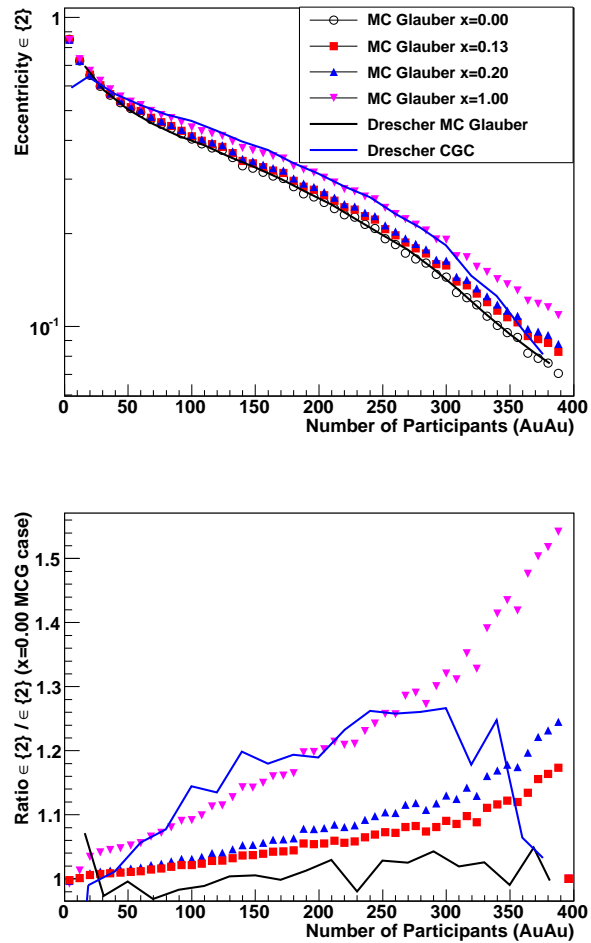}
\caption{\label{fig_eccentricity}Glauber Monte Carlo eccentricities.  Upper panel are the
values for $\epsilon_{part}\{2\}$ as a function of the number of participants (calculated with different assumptions).
Lower panel is the ratio of values relative to the Monte Carlo Glauber with $x$=0.0 case.}
\end{figure}

Fig.~\ref{fig_eccentricity} shows a new calculation of the eccentricity values using
a modified version of the PHOBOS MC code~\cite{phobos_glauber_hepforge} where we have
incorporated the two-component model.  Note that we use the standard nuclear parameters and 
no nucleon-nucleon hard core potential ($d=0$).
We calculate the eccentricity values from the Monte Carlo event-by-event
by the following equation:
\begin{equation}
\epsilon_{part} = {{\sqrt{(\sigma_{y}^{2}-\sigma_{x}^{2})^{2} + 4\sigma_{xy}^{2}}} \over {\sigma_{y}^{2} + \sigma_{x}^{2}}}
\label{eqn_eccentricity}
\end{equation}
The notation $\epsilon_{part}$ will be used even for $x \neq 0$, i.e. when the weighting is not just
for the participants.
For comparisons with PHOBOS data, the second cumulant
$\epsilon_{part}\{2\} = \sqrt{ \langle \epsilon^2_{part} \rangle }$ is used,
since the event plane method is effectively a two particle correlation~\cite{Alver:2008zza}.

Fig.~\ref{fig_eccentricity} shows the value of $\epsilon_{part}\{2\}$ for Au+Au collisions 
as a function of $N_{part}$ for four values of ${\it x}$ = 0.0, 0.13, 0.20, and 1.00.
The lower panel of Fig.~\ref{fig_eccentricity} shows the ratio of each case relative to the $x = 0.00$ case.
Results from Ref.~\cite{Drescher:2007cd} for ${\it x} = 0.20$ are also shown.
We find that our results for ${\it x} = 0.20$ are systematically larger than their results.  
This discrepancy has been traced to the fact that the
analysis in Ref.~\cite{Drescher:2007cd} uses a non-standard form
for Eqn.~\ref{eqn_twocomponent}, with a weight of one (as opposed to $1/2$) for the $N_{part}$ term.
This difference then gives a larger weighting for the spatial distribution of participants (i.e. lower ${\it x}$ value),
though there is no mapping onto a different exact ${\it x}$ value given by Eqn.~\ref{eqn_twocomponent}.
We also show the eccentricity values for their Color Glass Condensate calculation for comparison,
which interestingly track our ${\it x} = 1.00$ case, i.e. follow the density of the binary
collisions, except in the most central collisions.

We have also checked the eccentricity fluctuations for the different ${\it x}$ assumptions to see 
if this might offer an experimental method to discriminate between different $x$ values.  
The results show that the event-to-event
fluctuations in eccentricity are the same for all ${\it x}$ cases within 3-6\%.  
As one might expect, the fluctuations are somewhat smaller for the ${\it x}$ = 1.00 case,  
since there are a larger number of binary collisions to smooth out the 
fluctuations compared with participants.  However, this difference
is affected by the spatial correlations between binary collisions and might even be 
further modified with inclusion of a nucleon-nucleon hard core potential (i.e. a non-zero value 
of $d$).

In addition to the differences in eccentricities even for the case of identical ${\it x}$ values, there
are a number of differences we have uncovered that we describe below.

\begin{itemize}

\item{
The average overlap area $S_{T}$ is calculated
using the following equation:
\begin{equation}
S_{T} = 4 \pi \sqrt{\sigma_{x}^{2}\sigma_{y}^{2} - \sigma_{xy}^{2}}
\end{equation}
corresponding to the area of the tilted ellipse.
It should be noted that, in the literature, this equation appears in with various prefactors
(e.g. $\pi$ ~\cite{Alver:2008zza}, $2\pi$ ~\cite{Bhalerao:2005mm}, and $4\pi$~\cite{Gombeaud:2007ub}).
The prefactor $4 \pi$ gives a somewhat more intuitive result, 
as it yields the correct answer for the simple
case of a disc with uniform density.  This convention is utilized in the remainder of this paper.
However, since the parton density is not uniform, it is possible that
one should consider a smaller area (for example in a core/corona type picture).

In the extraction of the product $(\sigma c_{s})$ from
Eqn.~\ref{eqn_drescher}, any multiplicative scaling of the vertical-axis quantity does not change the
extracted results, but a multiplicative scaling of the horizontal-axis $\sim \frac{1}{S_T}\frac{dN}{dy}$ results in a linear change in $(\sigma c_{s})$.
However, we later note that simple multiplicative scaling of these quantities
cancels in the determination  $\eta/s$.

Finally, Ref.~\cite{Drescher:2007cd} does not calculate the overlap area $S_{T}$
in the tilted (i.e. fluctuating event-by-event) frame.  This turns out to be a minor (1-2\%) difference,
since the cross term $\sigma_{xy}$ turns out to be relatively small compared to the other
two moments.
}

\item{
The experiments do not measure the parton density at early times, but instead measure $dN_{ch}/d\eta$ 
after freeze-out.
The rapidity density $dN_{ch}/dy$ is extracted from the measured $dN_{ch}/d\eta$
by multiplying by a simple estimate of the Jacobian factor of 1.15.
However, this factor depends on the particle mix and momentum spectra, which are not measured
by PHOBOS over their full acceptance.
In the calculations of Drescher {\it et al.}, they have converted the $dN_{ch}/d\eta$ to
$dN_{ch}/dy$ with a factor 1.25; however, in our calculations we have used the factor
utilized by PHOBOS of 1.15\cite{Alver:2008zza}.  
One can then convert from charged hadrons to all hadrons by
assuming a multiplicative factor of 1.5.  It should be noted that equating this hadron
density to the {\it parton} density requires the additional assumption of local parton-hadron duality,
with an exact factor of 1.0 scaling.  Each of these factors enters as a linear scaling of the
horizontal-axis quantity $\sim \frac{1}{S_T}\frac{dN}{dy}$.
}

\item{
Ref.~\cite{Drescher:2007cd} calculates the average of the product
(i.e. $\langle 1/S_{T} \times dN/dy \rangle$) in
the Monte Carlo approachm, for both CGC and Glauber.  In contrast, since PHOBOS
uses the measured $\langle dN/dy \rangle$ and calculate $\langle S_T \rangle$ for each centrality
bin, they determine the product of averages (i.e. $\langle 1/S_{T} \rangle$ $\langle dN/dy \rangle$.).
If one wants to use the average of the product, one cannot directly utilize
the PHOBOS measured $dN_{ch}/d\eta$ as a function of $N_{part}$, since one no longer has a model
of the correlated fluctuations between the two quantities.
Instead one must use a parametrization based on the MCG calculation event-by-event.
Throughout this paper, we calculate the average of the product.   We have also checked in the MCG
calculation that taking the product of the averages and find differences in the extracted fit parameters of
order 20\%.

In our calculations, even though $\epsilon$ and $S_{T}$ are calculated with
different $x$ values, $x = 0.13$ is always used to determine the charged particle multiplicity.
This is because any case considered should have the correct particle multiplicity input (which is not
described by significantly different $x$ values within the two-component model).  Thus, one should
think of varying $x$ as an arbitrary way of modifying the initial geometry while maintaining the
correct multiplicity at a given centrality.  If one were certain that the
charged particle multiplicity constrains the geometry only around $x \approx$ 0.13, then these
other $x$ scenarios might be ruled out.
It is also notable that for the Cu+Cu data,
there is a disagreement between this parametrization of the charged particle multiplicity
at mid-rapidity by approximately 10\% in the most central events.
}

\end{itemize}


One final comment about the experimental uncertainties, and how they enter the fits.
The PHOBOS data are presented graphically with vertical lines for the quadrature sum of statistical
and systematic uncertainties at the 90\% Confidence Level.
Thus, they are scaled down by 1.6 to convert them into one standard deviation uncertainties
for input to a $\chi^{2}$ fit.
It is also not clear that the systematic uncertainties included here are
uncorrelated point-to-point, which raises some issues about
their inclusion in the standard $\chi^{2}$ fit.
Were the correlations known, the modified $\chi^{2}$ fitting procedure developed in~\cite{Adare:2008qa} could be used
to properly account for the correlations when determining the errors on the fitted parameters.
We make an estimate of this possible systematic uncertainty correlation in the next section.

\section{Extracting $\eta/s$}

\begin{figure}
\includegraphics[width=\columnwidth]{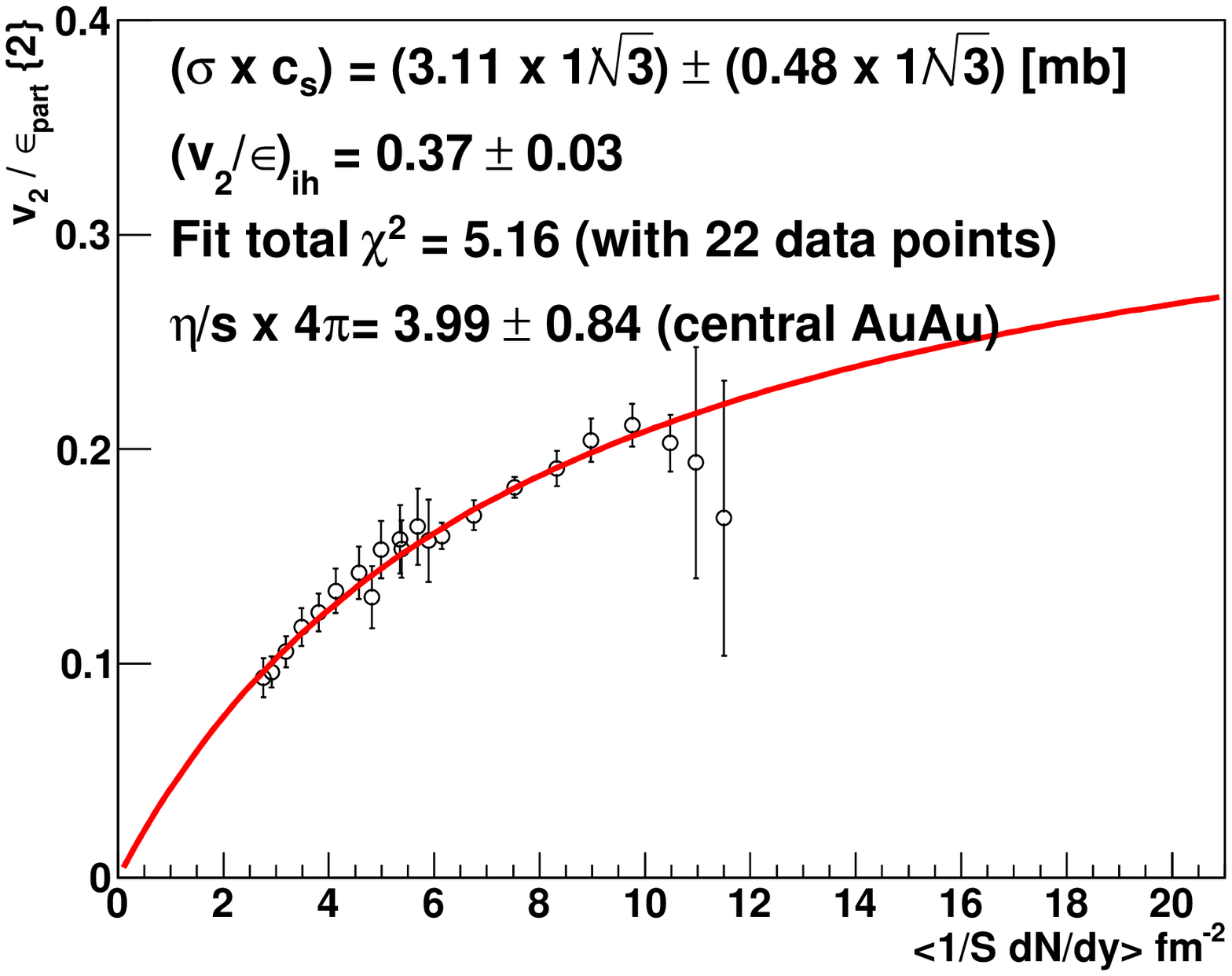}
\includegraphics[width=\columnwidth]{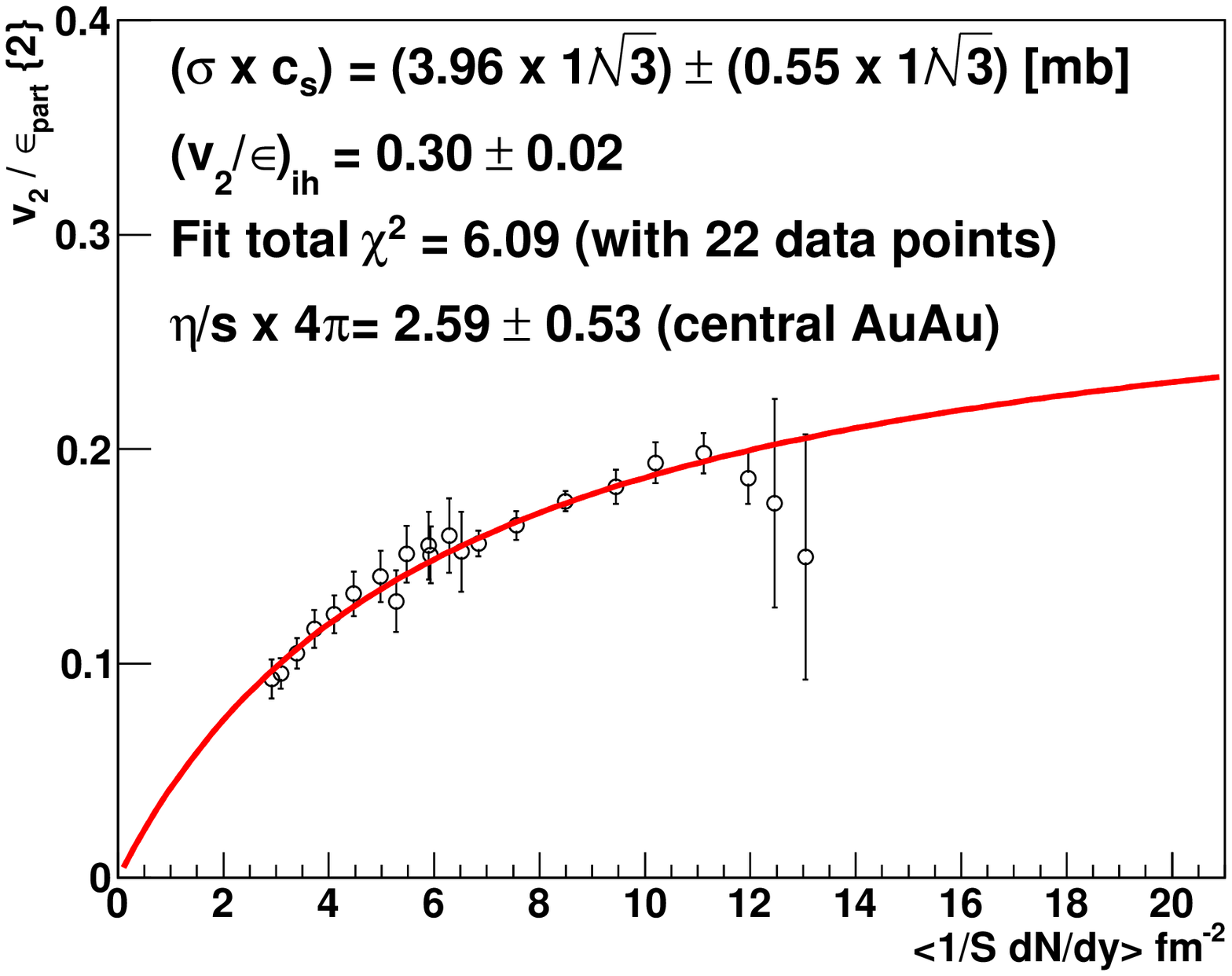}
\includegraphics[width=\columnwidth]{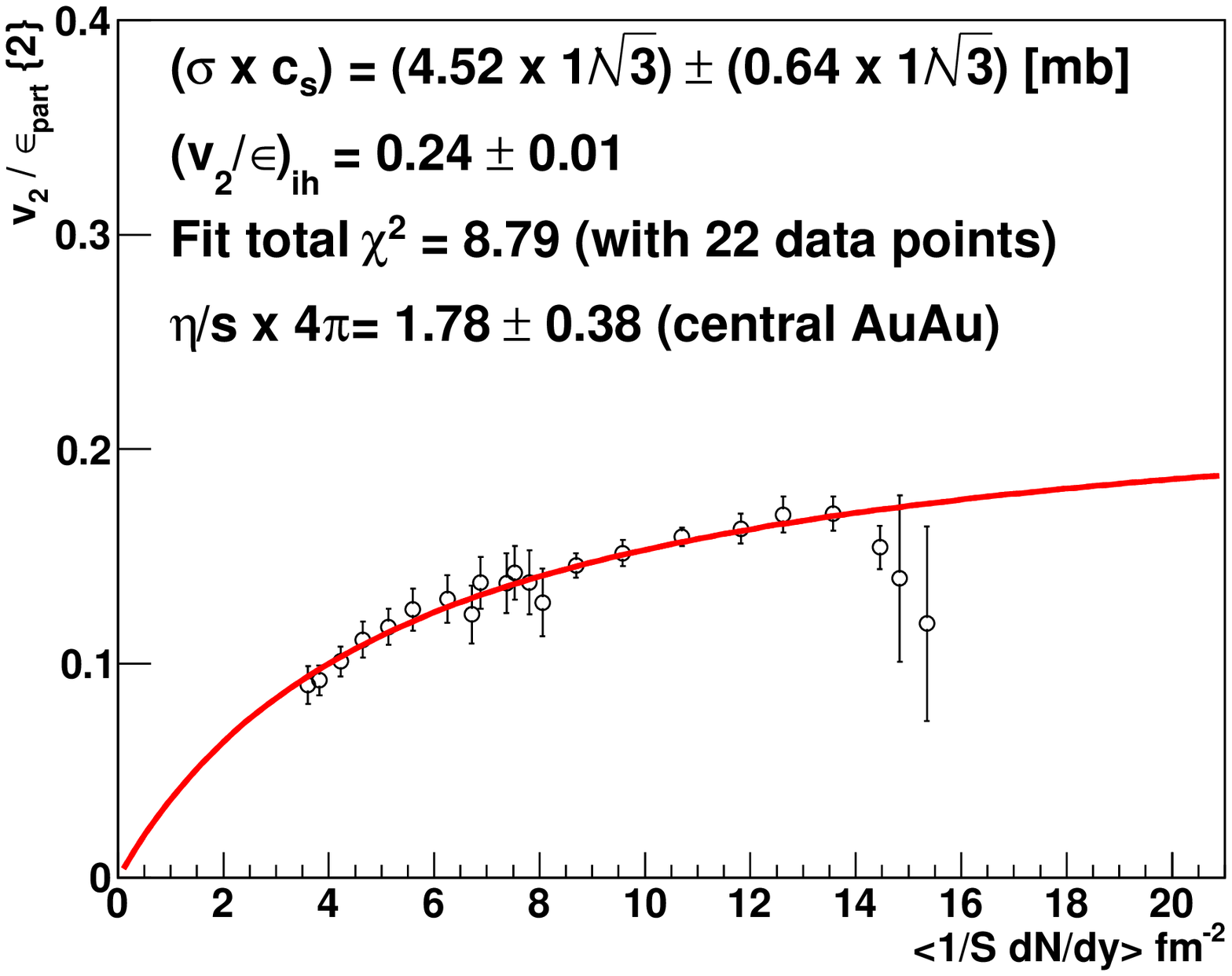}
\caption{\label{figmainfits}Shown are the PHOBOS data for $v_{2}/\epsilon_{part}\{2\}$
versus $\left< \frac{1}{S_T} \frac{dN}{dy} \right>\  \mathrm{fm}^{-2}$
where the $\epsilon_{part}\{2\}$ and $S_{T}$ are calculated from a Monte Carlo Glauber.  The
figures are using $x$=0.00, $x$=0.13, $x$=1.00 (in the top, middle, and bottom panel respectively).  
Also shown are the best fit results in each case for the two fit parameters ($\sigma \times c_{s}$) 
and ($v_{2}/\epsilon$)$_{ih}$.  Additionally,
following Eqn.~\ref{finaletas}, we calculate the $\eta/s \times 4\pi$ for each $x$ assumed case.
The quoted parameter uncertainties are one standard deviation fit uncertainties only,
assuming the experimental systematic uncertainties are point-to-point uncorrelated.
}
\end{figure}

Fig.~\ref{figmainfits} shows the results from our calculations following
the previously described prescription, using our
MCG results with $x$=0.00, $x$=0.13, $x$=1.00.
Given the experimental data for $v_2/\epsilon_{part}\{2\}$
as a function of transverse areal density, the fit procedure determines
values and uncertainties for ($\sigma c_{s}$) and $(\frac{v_2}{\epsilon})_{ih}$.
Since only the product $\sigma c_{s}$ appears in the fit function, determination
of the cross-section requires an explicit assumption for the sound speed.
Taking $c_{s} = 1/\sqrt{3}$ for definiteness  one obtains a cross-section value (in millibarns).
The extracted values are shown in Fig.~\ref{figmainfits}.

The extracted cross-section can be converted into a shear viscosity by using the results of a
classical calculation~\cite{Kox:76aa} for massless particles at temperature $T$ interacting with isotropic
cross-section $\sigma$
\begin{equation}
\eta = 1.264 \frac{T}{\sigma}
\end{equation}
giving
\begin{equation}
\frac{\eta}{s}
= 1.264 \frac{T}{s \sigma}
= 1.264 \frac{T}{4 n \sigma}
= 0.316 \frac{T}{\frac{(\sigma c_{s})}{\overline{R} S_{T}} \frac{dN}{dy} }
\label{eqn_deGrootForm}
\end{equation}
In the second step the entropy was assumed to be given by $s=4n$,
which follows from a definition of the number density using an ideal gas
equation of state $n = P/T$ and the assumption of massless quanta with
pressure $P=\frac{1}{3} \epsilon$ (here only $\epsilon$ is the energy density).
We note that there is already a 10\% ambiguity here since it is common practice in
astrophysics to compute the number densities directly from the Bose distribution
functions~\cite{KolbTurner}, leading to 3.6 units of entropy per photon.

Taking $T = 200$ MeV by assumption (another number with substantial uncertainty)
and values for  $\overline{R}$ from the Monte Carlo Glauber,
one can calculate $\eta/s$ for the most central Au+Au reactions at $\sqrt{s_{NN}} = 200$ GeV.  
The resulting values are shown in each panel of Fig.~\ref{figmainfits}, and
are summarized with later results in Fig.~\ref{etas_nsz_summary}.
It is notable that the initial conditions with larger eccentricity (e.g. ${\it x} = 1.00$) yield 
smaller values of $\eta/s$.  This is counterintuitive since a larger initial eccentricity should
require a larger viscosity to reduce the $v_{2}$ down to the experimentally measured value.  This
result highlights that the extracted value of $\eta/s$ comes from the curvature of the data
points as a function of centrality and the ${\it x} = 1.00$ case indicates the largest degree
of flattening (as seen in Fig.~\ref{figmainfits}).


In principle each parameter has its own physics content,
but it is interesting to check which parameters are necessary
in order to extract $\eta/s$ directly.
Since $K = n \sigma \overline{R}$, one can directly write:
\begin{equation}
\frac{\eta}{s} = 0.32 \frac{T}{n\sigma} = 0.32 K_{0} T \overline{R} \left[ \frac{(v_{2}/\epsilon)_{ih}}{v_{2}/\epsilon} - 1\right]
\label{finaletas}
\end{equation}
and quantities such as the speed of sound $c_{s}$ and the transverse overlap area $S_{T}$ do not explicitly appear.
As a result, this form has the considerable advantage of explicitly demonstrating the parametric dependence of
$\eta/s$ on the input parameters. Conversely, this simple expression holds only for the 'minimalist' assumption
for the $K$-dependence given by Eqn.~\ref{eqn_drescher}, and therefore imposes the need to investigate alternative
parametrizations of this dependence, as discussed in the following section.

It is interesting to compare our results with those from Ref.~\cite{Drescher:2007cd},
who obtain $\eta/s \times 4 \pi$ = 2.38 using Monte Carlo
Glauber ${\it x}$ = 0.20 (with the previously noted caveat regarding their non-standard ${\it x}$ definition) and $\eta/s \times 4 \pi$ = 1.38 using Color Glass Condensate initial conditions.
Before comparing these values to our results, we note that there is a mistake in their calculation of $\eta/s$ in the CGC case.  In the CGC case, in extracting
the cross-section $\sigma$, they assume a speed of sound $c_{s} = 0.73 \times 1/\sqrt{3}$, but then use a density $n$ calculated
with the input $c_{s} = 1/\sqrt{3}$.   As can be seen explicitly in our Eqn.~\ref{finaletas}, the speed of sound cancels out.  Thus,
the correct CGC value from their analysis is $\eta/s \times 4\pi$ = 1.89.  
We also note that their quoted value of $n = 3.9 \ \mathrm{fm}^{-3}$ from
reference~\cite{Bhalerao:2005mm} does not appear in that Letter, 
and must be converted for the difference in the $S_{T}$ prefactor, 
the use of optical Glauber versus Monte Carlo differences, 
and a missing Jacobian factor.

Taking these observations into account, we note that the values from our calculation for ${\it x}$ = 0.13 with $\eta/s \times 4\pi = 2.59 \pm 0.53$ for central Au+Au
is similar to that from Ref.~\cite{Drescher:2007cd} for their ${\it x}$ = 0.20 with $\eta/s \times 4\pi = 2.38$ (despite the
mismatch in exact eccentricities and other differences outlined above). Similarly, our calculation with ${\it x}$ = 1.00
gives $\eta/s \times 4\pi = 1.78 \pm 0.38$, similar to Ref.~\cite{Drescher:2007cd} for the CGC case with
$\eta/s \times 4\pi = 1.89$ (with the correction described above).

Here we re-visit the issue of the handling of the PHOBOS experimental systematic uncertainties.  
In the above we have considered them to be point-to-point uncorrelated.  Since we do not
know the full correlation matrix, we consider two cases to estimate of the systematic uncertainty
on the extracted $\eta/s$.  If we allow the data points to move within the one standard deviation
uncertainties in a correlated or anti-correlated manner (i.e. tilting the $v_2 / \epsilon_{part}\{2\}$ values
around the mid-central point), we find that for the ${\it x}$ = 0.13 case, we obtain values of 
$\eta/s = 1.68$ (moving the central points down and the peripheral points up) and $\eta/s = 3.61$ 
(moving the central points up and the peripheral points down).  We note that the $\chi^{2}$ total 
is very large in all cases in part because the statistical uncertainties are quite small and some
part of the systematic uncertainty is likely uncorrelated.  Thus, as an approximate estimate, the
true fit uncertainty from the experimental uncertainties alone is most likely twice as large
as that shown in Fig.~\ref{figmainfits}.  For the ${\it x}$ = 0.13 case, we then should state it
as $\eta/s = 2.59 \pm 1.00$, instead of the previously quoted $\eta/s$ = $2.59 \pm 0.53$.

\section{Other Parametrizations}
It has already been noted that Eqn.~\ref{eqn_drescher} is not derived from any
{\it a priori} theoretical expectation,
but has been constructed to obey limiting functional forms at large and small $K$.
Given this, it is of interest whether an alternative functional form might fit
the available data equally well and still obey the same two limits.  A simple
functional form (motivated by Pad\'e approximants) is utilized here:

\begin{equation}
\frac{\frac{v_{2}}{\epsilon}}
{\left( \frac{v_{2}}{\epsilon} \right)_{ih}}
=
\frac{1 + A  \frac{K}{K_{0}} + B \left( \frac{K}{K_{0}} \right)^2 }
{1 + C \frac{K}{K_{0}} + D \left( \frac{K}{K_{0}} \right)^2 + E \left( \frac{K}{K_{0}} \right)^3}
\end{equation}

\begin{figure}
\includegraphics[width=\columnwidth]{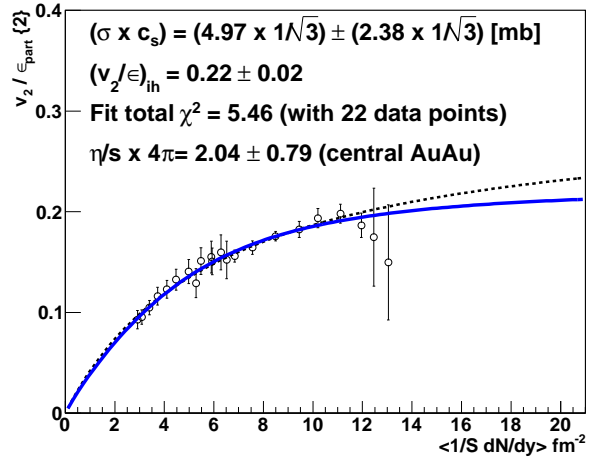}
\caption{\label{figure_padetypefits} Shown are the PHOBOS data for $v_{2}/\epsilon_{part}\{2\}$
versus $\left< \frac{1}{S_T} \frac{dN}{dy} \right>\ \mathrm{fm}^{-2}$
where the eccentricity and transverse overlap area are calculated from a Monte Carlo Glauber.  The
calculation uses $x$=0.13.  Also shown are the best
fit results (solid line) using the formulation with the Pad\'e approximant motivated functional form, and then extracting the fit
parameters ($\sigma \times c_{s}$) and $v_{2}/\epsilon_{ih}$).
Additionally, following Eqn.~\ref{eqn_deGrootForm}, we calculate the $\eta/s \times 4\pi$.  For comparison, we plot
the curve for the same case with the standard fit (dashed line).
}

\end{figure}

This equation obeys the large and small $K$ limits as the previous parametrization,
if  $B=E$ and $C = A + 1$.
Of course one can obtain an infinite set of such equations by expanding
the number of terms in the numerator and denominator.  For this case, we
constrain ourselves to the case where $B = E = 1$ and $A = 1, C = 2$.
Fig.~\ref{figure_padetypefits} shows the best fit where the
additional parameter $D$ is allowed to vary (for the ${\it x}$ = 0.13 case).
There is an approximately 25\% decrease in the ideal hydrodynamic value
$(\frac{v_{2}}{\epsilon})_{ih}$, while the $\sigma = 4.97 \pm 2.38$
millibarns (assuming $c_{s} = 1/\sqrt{3}$) increases by about 25\%
relative to the previous value of $3.96 \pm 0.55$, although with a larger uncertainty.
This best fit corresponds to $D \approx 2$.

\begin{figure}
\includegraphics[width=\columnwidth]{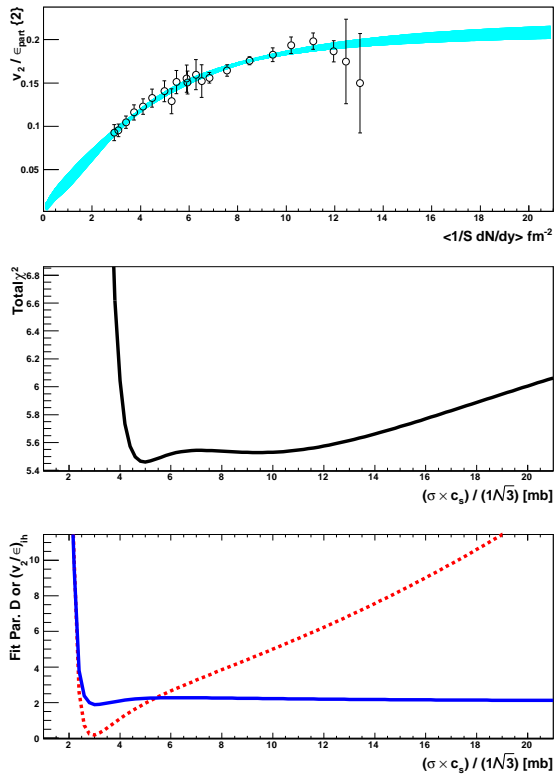}
\caption{\label{pade2}
Upper panel shows the fit results with $\chi^{2}$ values within
$1\sigma$ of the minimum.  Middle panel shows the total $\chi^{2}$ as a function
of the extracted $\sigma c_{s}$ value.  Lower panel shows the associated parameter values
for $D$ (dashed) and $\frac{v_{2}}{\epsilon}_{ih}$ $\times 10$ for visibility (solid).
}
\end{figure}

However, it turns out that the standard MINUIT $\chi^{2}$ fit has returned
an incorrect uncertainty range because of a
non-parabolic shape of the $\chi^{2}$ surface.
The reason for this behavior is examined in
Fig.~\ref{pade2}, which shows all of the parameter fits with $\chi^{2}$ values within
one standard deviation of the minimum, the total $\chi^{2}$ as a function
of the extracted $\sigma c_{s}$ value, and the associated parameter values
for $D$ and $\frac{v_{2}}{\epsilon}_{ih}$.

The presented $\chi^2$ as a function of ($\sigma c_s$) allows the determination of
the uncertainty on $\sigma$ at one-standard deviation to be $^{+25.0}_{-1.0}$ millibarns.  
Almost any arbitrarily large value for the cross-section $\sigma$ gives an equally
good description of the experimental data with an almost identical value for
$\frac{v_{2}}{\epsilon}_{ih}$ and an increasing value of $D$ for larger $\sigma$
values.
Using Eqn.~\ref{eqn_deGrootForm} to determine the corresponding values for
$\eta/s$ we find an allowed range of $\eta/s \times 4\pi$ from 2.55
all the way down 0.34 (well below the postulated bound).

These results indicate a certain fragility in the procedure outlined in Ref.~\cite{Drescher:2007cd} 
since a minimal modification of the parametrization
amplifies the uncertainties in $\eta/s$ by an order of magnitude.
It is therefore important to investigate if there are {\it a priori} theoretical restrictions
on the functional dependence of $v_2/\epsilon$ on the Knudsen number.
For example, are there theoretical arguments that might constrain $D \sim 2$.
In a simple transport scenario with centrality independent cross-section $\sigma$,
temperature $T$, and speed of sound $c_{s}$, one might question why the underlying dynamics
would change significantly with centrality (i.e. $D >> 1$).

However, if obeying the two limits for $K$ is insufficient to determine the functional
form, then one needs to make a proper theoretical argument
about which types functional forms are truly relevant.  There have been attempts to check the
Knudsen number dependence with transport calculations and viscous hydrodynamic calculations~\cite{Masui:2009pw}, 
and these may lend support to the simple parametrization.  However, if the parametrization is only
tested in certain scenarios, then one has all the limitations 
of those particular scenarios and one cannot make
more general conclusions.

\section{Quantum Limits}

One obvious feature of the above treatment is the lack of any explicit quantum bound on
the value for $\eta/s$.  Clearly if such a bound exists, then the experimental values for $v_2/\epsilon$
cannot approach the ideal hydrodynamic limit, and it is unlikely that a transport formalism that allows
violations of the bound would be applicable to the determination of $\eta/s$ in the vicinity of the bound.
In this section we consider a modification of the formulation
which explicitly incorporates the bound.

Following the original formulation of the bound~\cite{Danielewicz:1984ww}, we assume that the
mean free path $\lambda$ cannot be smaller than the DeBroglie wavelength of the particle.
A simple prescription incorporating this limit is to modify the mean free path accordingly:
\begin{equation}
\lambda = \frac{1}{n \sigma} \longrightarrow \frac{1}{n \sigma} \left[ \frac{1}{1-e^{-\langle p \rangle/n\sigma}} \right]
\end{equation}
\noindent
In the limit where $\langle p \rangle \ll n \sigma$, $\lambda = 1/\langle p \rangle$
(the DeBroglie wavelength), while
the limit $\langle p \rangle \gg n \sigma$ gives the standard expression $\lambda = 1/ (n\sigma)$.
Note that we are using natural units throughout.

In relating these values to $\eta/s$ we proceed as before
\begin{equation}
\frac{\eta}{s} = 0.316 \frac{T}{n \sigma} \longrightarrow 0.316
\frac{T}{n\sigma{\left[ 1-e^{-\langle p \rangle/n\sigma} \right]}}
\end{equation}
Taking  $\langle p \rangle \approx 2.7 T$
(relevant for massless, non-interacting particles)
the high density limit gives 
$\eta/s \approx 0.316 \ T \times \lambda = 0.316 T / (2.7 T) = \frac{1.4}{4\pi}$.
Thus, there is a modest inconsistency (at the 40\% level) between
this implementation and the exact bound value.  Note that  it is possible to
enforce the precise value of the bound by requiring $\lambda \geq 0.7/\langle p \rangle$,
but in the interest of a simpler heuristic treatment we do not consider this additional modification.
We highlight that even in the presence of quantum effects the
Knudsen number itself $K = \lambda / \overline{R}$ can come arbitrarily close to zero,
for example in a neutron star where $\overline{R} \gg $ any other
scale in the problem.  

The expression for $K$ as a function of $\frac{1}{S_T} \frac{dN}{dy}$ is:
\begin{equation}
K = \frac{\lambda}{\overline{R}} = \frac{1}{\overline{R}} \frac{1}{n \sigma} \left[ \frac{1}{1-e^{-2.7T/n\sigma}} \right]
\end{equation}
\begin{equation}
K = \frac{1}{\sigma c_{s} \langle \frac{1}{S_{T}}\frac{dN}{dy} \rangle}
\left[\frac{1}{1-e^{-2.7 T \overline{R}/(\sigma c_{s} \langle \frac{1}{S_{T}}\frac{dN}{dy} \rangle)}}\right]
\end{equation}
However, when using this expression into
Eqn.~\ref{eqn_drescher} one needs to know the
dependence of $\overline{R}$ on $\left< \frac{1}{S_T} \frac{dN}{dy} \right>$ before performing the fit.
This dependence has been calculated with a Monte Carlo Glauber and
then included in the fit to the experimental data.

\begin{figure}
\includegraphics[width=\columnwidth]{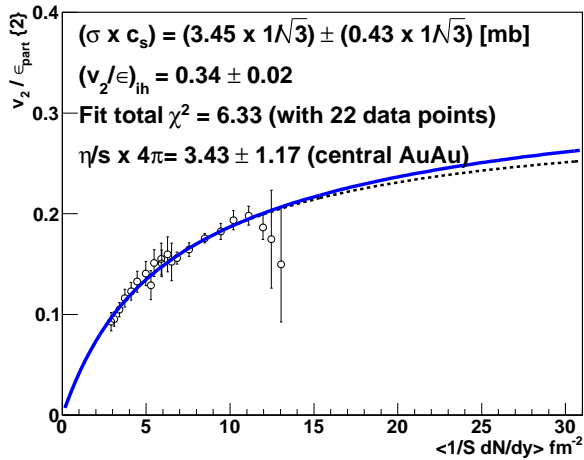}
\caption{\label{figure_quantumcase} Shown are the PHOBOS data for $v_{2}/\epsilon_{part}\{2\}$ versus
$\left< \frac{1}{S_T}\frac{dN}{dy} \right>\ \mathrm{fm}^{-2}$
where the eccentricity and transverse overlap area are calculated from a Monte Carlo Glauber.  The
calculation uses $x$=0.13.  Also shown are the best
fit results using the formulation with the quantum bound correction limit (solid line), and then extracting the two fit
parameters ($\sigma \times c_{s}$) and $(v_{2}/\epsilon)_{ih}$.  Additionally,
following Eqn.~\ref{eqn_deGrootForm}, we calculate the $\eta/s \times 4\pi$.  For comparison, we plot
the curve for the same case without the quantum bound correction (dashed line).
}
\end{figure}

The resulting value of the fit is shown in Fig.~\ref{figure_quantumcase}.
It gives a slightly smaller value of $\sigma c_{s} = 3.45 \pm 0.43 (\times 1/\sqrt{3})$ millibarns.
This modest change is not surprising since the starting
value without the correction is already a factor of 2.5 away from the quantum bound.
For systems even closer to the bound, the impact would be much more significant.
Note however that the extracted value of $\eta/s$ is almost $1/4\pi$ larger than that extracted without
incorporating the bound.  This is not at all unexpected since one has re-interpreted the associated
$\eta/s$ value with the asymptotic limit of the fit function.  


\section{Discussion}

\begin{figure}
\includegraphics[width=\columnwidth]{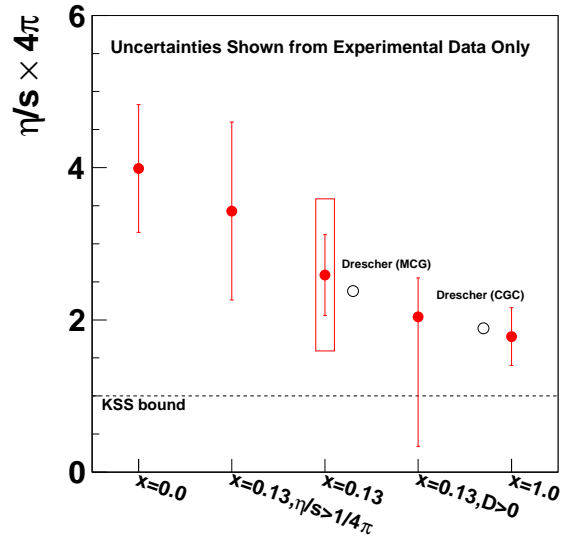}
\caption{\label{etas_nsz_summary}The solid points are the extracted values of $\eta/s$ as a function
of Monte Carlo Glauber parameter ${\it x}$ and for the quantum limit and modified Knudsen number
parametrization cases.  The vertical uncertainty lines are the one standard deviation uncertainties from
the fits to the experimental data assuming the PHOBOS systematic uncertainties are point-to-point
uncorrelated.  An estimate of the systematic uncertainty on $\eta/s$, if the experimental
systematic uncertainties are correlated (as discussed in the text),
is shown as a box for the ${\it x} = 0.13$ case.  Also shown as open symbols are the
values extracted by Drescher {\it et al.} for Glauber and CGC initial conditions.
}
\end{figure}

In this paper, we are not advocating the Knudsen number formalism as a 
rigorous method of extracting $\eta/s$.  Rather
we are highlighting various issues with this  methodology and note areas 
where systematic uncertainties are
common to various methods that have been proposed for extracting $\eta/s$.  
There are three main conclusions.

1.  As has been previously observed, any method for extracting $\eta/s$ 
will be sensitive to the details of the initial conditions,
and in particular the spatial distribution (and fluctuations) 
of the deposited energy density.
This induces a strong sensitivity to the assumed ratio $x$ of 
binary collisions to participants in
the Monte Carlo Glauber formalism since a modest centrality dependent
change in the spatial distribution substantially changes the asymptotic 
behavior implied by the fit.  This is particularly true
because the experimental data does not have small enough uncertainties 
for central Au+Au events to determine whether an asymptotic limit has been reached.  
This is the main reason why choosing a different functional form for the
Knudsen number parametrization allows almost any value for $\eta/s$ to 
be extracted, as demonstrated above.

In this paper we do not consider the Color Glass Condensate initial conditions 
in detail.  However we note that there are at least two formulations that 
give significantly different results.
These include calculations such as Ref.~\cite{Lappi:2006xc}
which do not include fluctuations, and Ref. \cite{Drescher:2006ca}
which includes fluctuations and is the basis of 
the calculations by Drescher {\it et al.} shown in Fig.~\ref{fig_eccentricity}.  
The differences in CGC initial condition eccentricities are as large as the
differences between Monte Carlo Glauber ${\it x}$ values, and thus this also needs
to be addressed and reconciled.


These initial geometry uncertainties generally have a larger influence on this extraction method since 
a subtle difference in centrality dependence (i.e. curvature) can
dramatically change the fit.
Note that while Ref.~\cite{Alver:2008zza} 
finds the change in eccentricity from $x=0.0$ to $x=0.13$ for central Au+Au events to be 15\% and
thus considered small, we find that the change in the centrality dependence modifies
the extracted value of $\eta/s$ by more than 50\%.

These uncertainties certainly also influence other methods of extracting $\eta/s$.
We note that in the viscous hydrodynamic calculations in Ref.~\cite{Luzum:2008cw}, they consider 
two sets of initial conditions.  In the Glauber case they utilize an optical model without fluctuations, and 
utilize the ${\it x}$ = 1.00 case.  In contrast, Ref.~\cite{Song:2007ux} utilizes participants for
initial conditions (i.e. ${\it x} = 0.00$).
In the case of constraints on $\eta/s$ from charm and bottom elliptic flow 
$v_2$~\cite{Adare:2006nq} the extractions of $\eta/s$ will also depend 
on the initial conditions and the same scrutiny and propagation of uncertainties is needed.
In Ref.~\cite{Moore:2004tg} they utilize an optical Glauber model without fluctuations and wounded
nucleons (i.e. ${\it x} = 0.00$), while in Ref.~\cite{vanHees:2005wb} they state that for impact parameter $b=7$ fm
the eccentricity is $\epsilon = 0.6$.  

These observations highlight the critical importance of a reproducible set 
of publicly available code for generating what are
often simply referred to as ``Glauber'' initial conditions.  All 
publications need a detailed specification of all relevant features of the
Glauber calculation, including
1) the $x$ value,
2) Glauber parameters for the Woods-Saxon distribution,
3) any $d$ exclusion radius and re-weighting of the radius value,
4) any outer radius cut-off,
5) whether fluctuations in the reaction plane direction are included (as seen in Eqn.~\ref{eqn_eccentricity})
6) whether the result is not a Monte Carlo but instead an optical Glauber extraction.  
Many of the individual uncertainties related to these parameters have previously been considered 
minor (i.e. 20\% or less).  

2.  Even if one accepts the assumptions of the formalism from Ref.~\cite{Drescher:2007cd}, 
there are significant uncertainties in the parameters characterizing the medium on time scales corresponding
to the largest gradients, i.e. $T$, $c_{s}$ and $\bar{R}$.  
Since these values vary as a function of both the space and time coordinates, one needs to specify the precise
time and space interval over which the averaging is performed.  It is also necessary to demonstate that such averaging of intermediate quantities is handled in a self-consistent fashion when deriving other
averaged quantities
(such as $\eta/s$).  For example, the speed of sound $c_s$ is introduced into Eqn.~\ref{eqndensity} when estimating the characteristic time over which
gradients develop.  Absent a detailed simulation, the speed of sound in such an expression is an ill-determined average over times prior to the desired characteristic time.  Nor is it obvious that 
this is same time-average assumed for other quantities such as 
$T$, $\bar{R}$, $\lambda$, etc.  




Moreover, there is also a potential inconsistency with the parameters used in studies up to
this point.  The temperature, parton density,
speed of sound, and interaction cross-sections are not independent parameters.
In fact at zero baryon density, the temperature determines all of the others.  
As an explicit example, a temperature of $T = 200$ MeV is inconsistent with 
the speed of sound of $1/\sqrt{3}$ when comparing with lattice QCD results at zero baryon 
chemical potential~\cite{Karsch:2008fe}.
Also, temperature and parton density can be translated into 
$T$ and $\epsilon$ (energy density) via the QCD equation of state.
While it seems reasonable that the temperature, density and speed of sound
are approximately constant over the range of densities for Cu+Cu and Au+Au 
collisions at a single colliding energy (e.g. $\sqrt{s_{NN}} = 200 $ GeV) 
considered here,
this is not expected to be true at asymptotically high densities.
Even though there are no experimental data points at such densities, 
and one extracts $\sigma \times c_{s}$ and $\eta/s$ at the density corresponding 
to Au+Au central collisions, even small changes can systematically alter the
centrality dependence, modifying the extracted value of $K$, and thus $\eta/s$.
One must also consider the possibility of temperature dependent cross-sections
in medium, something not addressed so far.  Again, what are often dismissed as
small centrality dependencies can modify the curvature of the data and have an
unexpectedly large impact on $\eta/s$.

We note that there has been a recent attempt to calibrate the simple analytic approach
discussed here, utilizing viscous hydrodynamics calculations~\cite{Masui:2009pw}.
These studies have already revealed an ambiguity in the value of 
$\eta/s$ and the equation of state.  Additionally, any such calibration
will have the same caveats and limitations of the existing viscous 
hydrodynamic calculations (with one such example being the absence of a hadronic phase after freeze-out).

3.  
In the formulation discussed in this paper, the extracted value of $\eta/s$ 
depends on whether or not a lower bound is explicitly introduced.
Thus, if a lower bound on the viscosity does in fact exist,
then $\eta/s$ extractions which do not incorporate it in their formalism will
ultimately be unreliable. 
Conversely, approaches which enforce such a quantum bound on $\eta/s$ 
obviously cannot extract values below the bound.   Thus one could never falsify the
existence of such a bound.
This is in contrast to full viscous hydrodynamic calculations and comparisons 
which do not rely on microscopic modeling of transport phenomena.

This raises the question at to whether the calculational framework discussed in
Drescher {\it et al.}, and this work, is valid in the strong-coupling limit, i.e. at
$\eta/s=1/4\pi$. 
A standard Boltzmann transport calculation is the opposite limit from hydrodynamic 
calculation, which assume a continuum hypothesis.  The theory community has explored both
approaches with great benefit, and in both cases one can attempt to push the models past
their nominal domains of applicability.  
If the viscous corrections to the ideal hydrodynamics
are too large, then they violate the gradient expansion made in the small $\eta/s$ limit.  
In the Boltzmann transport approach (e.g. a parton cascade),
the quasi-particle widths become of same order as the masses
when the mean free paths are comparable to the DeBroglie wavelengths
(i.e. $\Gamma/m \approx 1$)~\cite{LindenLevy:2007gd}.   
As an example, if one ignored multi-particle quantum effects in a gas of cold atoms near the 
BEC transition that would have a dramatic impact on the physics picture.
Thus, in this case, it is not clear how to determine the systematic uncertainties 
and overall physical picture.

\section{Summary}
The formalism for relating $\eta/s$ to the  Knudsen number 
has attracted much recent attention since it offers a tantalizing possibility
of extracting transport properties almost directly from experimental data. 
However, several important issues need to be addressed consistently 
before one can consider this approach to produce anything more than order-of-magnitude estimates.
The linearity of $\eta/s$ with $T$, $c_s$, and even $\bar{R}$ makes a serious evaluation of their
uncertainties of utmost important.  One must also consider the fact that $c_s$ is a function
of temperature due to the QCD equation of state itself.
The final result for $\eta/s$ is quite sensitive to Monte Carlo Glauber parameters, and particularly the
parameter $x$ which controls the balance between $N_{part}$ and $N_{coll}$.  
Finally, implementing a lower bound to $\eta/s$, for instance as we have presented in this work, has a
substantial effect on the extracted value, particularly since previous calculations find values so
close to the bound already.

\begin{acknowledgments}
We gratefully acknowledge useful discussions with Adrian Dmitru, Jean-Yves Ollitrault, Paul Romatschke, and Paul Stankus.
JLN acknowledges support from the United States Department of Energy Division of Nuclear Physics grant DE-FG02-00ER41152.
PAS is supported by U.S. Department of Energy grant DE-AC02-98CH10886.
WAZ is supported by U.S. Department of Energy grant DE-FG02-86ER40281.
\end{acknowledgments}
\bibliographystyle{h-physrev3}
\bibliography{nsz_knudsen}

\end{document}